\documentclass[%
reprint,
amsmath,amssymb,
aps,
showkeys
]{revtex4-2}

\usepackage{graphicx}
\usepackage{dcolumn}
\usepackage{bm}
\usepackage[pdfborder={0 0 0}]{hyperref}

\usepackage{physics}
\usepackage{overpic}
\usepackage{capt-of}

\usepackage{xcolor}
\newcommand{\stephan}[1]{{\color{green}#1}}

\usepackage{comment}

\begin{document}
	
	\preprint{APS/123-QED}
	
	\title{Quantum repeater node with free-space coupled trapped ions}
	
	\author{Max Bergerhoff}
	\thanks{These authors contributed equally to this work.}
	\author{Omar Elshehy}
	\thanks{These authors contributed equally to this work.}
	\author{Stephan Kucera}%
	\author{Matthias Kreis}%
	\author{J\"urgen Eschner}%
	\email{juergen.eschner@physik.uni-saarland.de}
	\affiliation{%
		Fachrichtung Physik, Universit\"at des Saarlandes, Campus E2.6, 66123 Saarbr\"ucken, Germany 
	}%

	\date{\today}
	
	\begin{abstract}
		The quantum repeater cell is a basic building block for a quantum network, as it allows to overcome the distance limitations due to unavoidable fiber loss in direct transmission. We demonstrate the implementation of a quantum repeater cell, based on two free-space coupled $^{40}$Ca$^+$ ions in the same trap that act as quantum memories. We demonstrate the asynchronous generation of atom-photon and photon-photon entanglement by controlled emission of single photons from the individually addressed ions and entanglement swapping. We discuss the fidelity as well as the scaling of the generated rate.
	\end{abstract}
	
	
	\maketitle
	
	\section{Introduction}
	Large quantum networks \cite{Quantuminternet, QuantumInternet_AVision} with single photons as flying qubits require tools for overcoming propagation loss. To this end, quantum repeaters \cite{Briegel_Quantenrepeater, Quantum_repeaters_overview, Azuma_2023, Avis_2023} using quantum memories and entanglement have been proposed. The fact that quantum signals cannot be amplified or copied because of the no-cloning theorem \cite{No-cloning_theorem} is then counteracted by distributing entanglement over smaller distances. This enables networking applications such as qubit transmission by quantum teleportation \cite{Bennett1993}, or quantum key distribution (QKD) \cite{BB84}, or distributed quantum gates \cite{Gottesman1999}. Two basic building blocks of a quantum repeater link are identified  according to \cite{whitepaper}: the quantum repeater segment, where two memory qubits are connected via a photonic link, and the quantum repeater cell (QR cell) where two closely spaced and interacting memory qubits are each connected to a photonic link. Using this configuration, the key generation rate for QKD is proven to be advantageous compared to the direct link \cite{Luetkenhausprotokoll}. 
	
	Concatenation of QR cells and segments enables entanglement distribution over an arbitrary distance. This employs asynchronous generation of memory-photon entanglement, quantum gates on the memories, and photonic Bell measurements. Important characteristics are, therefore, the indistinguishability of photons sent out by the memories, the fidelity of the atom-photon entanglement, and the coherence time of the memories. Furthermore, high rates are desirable for practical network applications.
	
	Implementations towards a quantum repeater are being developed on different platforms, such as vacancy centers \cite{bhaskar_2020, Stolk_2022}, atomic ensembles \cite{atoms,Liu2024}, single atoms coupled in free space \cite{vanLeent_2022} or in a cavity \cite{PhysRevLett_Rempe}, and ions in a macroscopic cavity \cite{Ion_Innsbruck}. Realizations of a QR cell that include the basic demonstration of the repeater advantage have so far been achieved with atoms in a cavity \cite{PhysRevLett_Rempe} and ions in a macroscopic cavity \cite{Ion_Innsbruck}. Trapped-ion based quantum memories are promising candidates for realizing quantum repeater protocols \cite{PhysRevA_ion_QR, QR_APE}, as the necessary elements for their control are well developed and can be combined with relatively low effort. This platform offers long coherence times \cite{Schmidt-Kaler_2003, Wang_2021}, the possibility to generate high-fidelity memory-photon entanglement \cite{Duan_2004, APE} and coherent manipulations to implement two-ion quantum gates and a Bell-state measurement between the memory qubits for entanglement swapping \cite{ion_entanglement_swapping, Ballance_2016}.
	
	In this manuscript, an implementation of a QR cell is demonstrated, based on two $^{40}$Ca$^+$ ions in the same trap that act as quantum memories, coupled to photonic channels in free space. In \autoref{sec:Experiment} an overview on the experimental setup and protocols is given. The core element of the protocol is the asynchronous generation of atom-photon entanglement which is described in detail in \autoref{sec:ExperimentATPh}. The generated photons from each atom are free-space coupled into separate single mode fibers. 
	The entanglement swapping from the memory to the photons, in order to generate the targeted photon-photon entanglement, is described in \autoref{sec:ExperimentPhPh}. It is implemented by the use of the M\o{}lmer-S\o{}rensen quantum gate \cite{MSOriginal,MS2} and subsequent projective measurement of the atomic states. In \autoref{sec:Results} the experimental realization is presented. First the asynchronous generation of atom-photon entanglement is demonstrated (\autoref{sec: ExperimentATPhResults}), then the state of the two photons after applying the entanglement swapping procedure is characterized (\autoref{sec: ExperimentPhPhResults}). Finally the scaling of the coincidence rate and probability is discussed (\autoref{sec: ExperimentScalingResults}).
	
	The experiment presented in the recent publication \cite{Ion_Innsbruck} follows a similar approach, also using single $^{40}$Ca$^+$ ions as quantum memories. The main differences are in some implementation steps of the protocol and, importantly, in the way of collecting the photons. While in \cite{Ion_Innsbruck} individually addressed excitation of the ions and photon collection via a large optical cavity are used, in our experiment individually addressed repumping is performed, and photon collection happens into free space. In the final discussion we will compare the rates of entangled photon pairs achievable with the two approaches.

	\section{Experiment}\label{sec:Experiment}
	
\begin{figure}[h]
		\centering
		\includegraphics[scale=0.65]{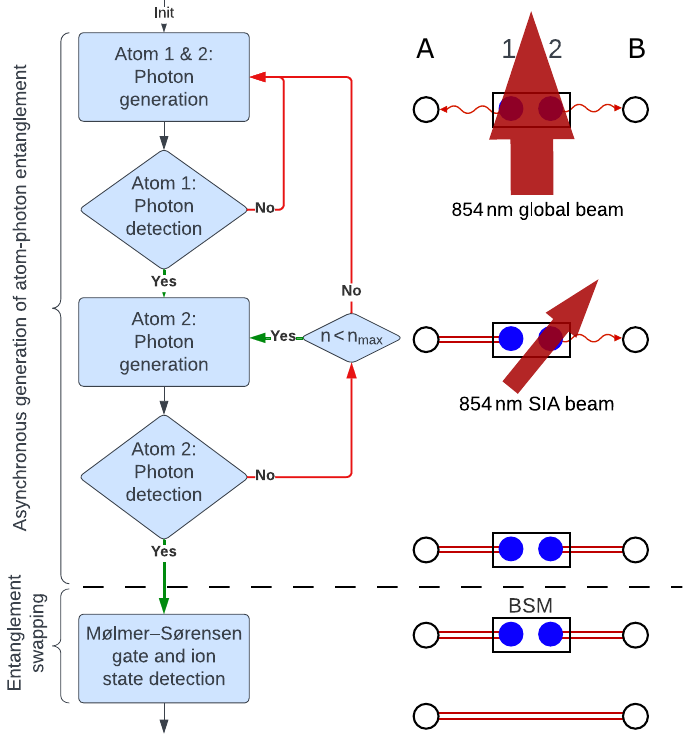}
		\caption{Sequence for asynchronous generation of photon-photon entanglement. For each step of the sequence on the left side, the right side shows the corresponding action on the two memory atoms (blue circles) to generate photons (black rings). The wavy arrows represent the emission of photons, and the straight lines represent entanglement. (SIA = single-ion-addressing; BSM = Bell state measurement)}\label{fig:seq_protocol}
	\end{figure}

	The experimental protocol for generating the targeted photon-photon entanglement is schematically shown in \autoref{fig:seq_protocol}. 
	The protocol is divided into two parts: the first part creates atom-photon entanglement between atom\,1 and a photon in arm A, and asynchronously between atom\,2 and a photon in arm B. A detailed description is given in \autoref{sec:ExperimentATPh}. The second part performs the entanglement swapping from the atoms to the photon pair, by projecting the two atoms onto a basis of maximally entangled states (Bell-state measurement, abbreviated as BSM). Details are explained in \autoref{sec:ExperimentPhPh}. 
	
\begin{figure*}[!]
		\centering 
		\includegraphics[scale =0.9]{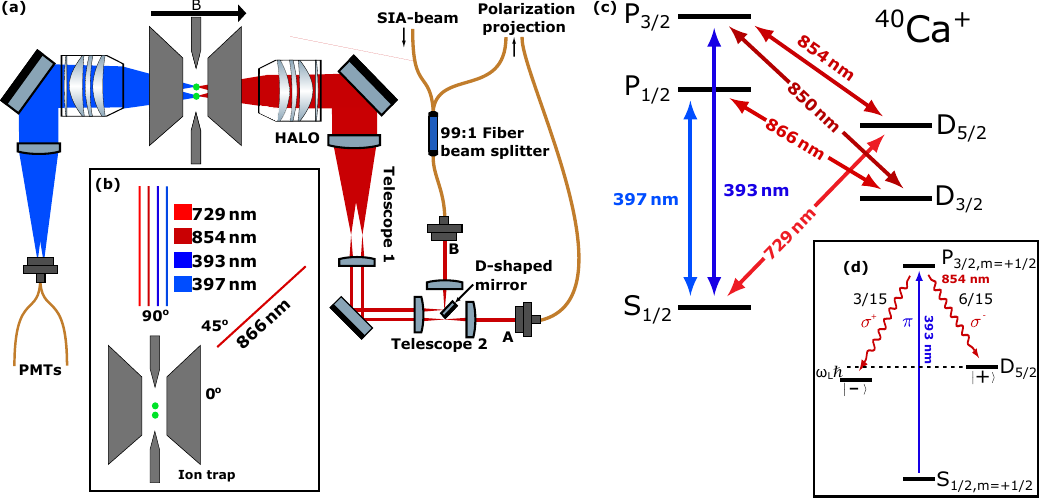}
		\caption{Schematic of the experimental setup. (a) Ion trap with HALOs and optical paths for single-atom-addressing, two-photon collection, and detection. (b) Geometric arrangement of laser beams used in the experiment. (c) Level scheme of the $^{40}$Ca$^{+}$ ion. (d) Zeeman sub-levels for generation of atom-photon entanglement, showing excitation and emission paths. Only the relevant levels for generation of the target state (\autoref{eq: ape}) are shown. Excitation of the initial state $\ket{\text{S}_{1/2},+1/2}$ to $\ket{\text{P}_{1/2},+1/2}$ with $\pi$-polarized 393\,nm light triggers decay to a superposition of $\ket{-}=\ket{\text{D}_{5/2},-1/2}$, with amplitude (Clebsch–Gordan coefficient) $\sqrt{3/15}$ and $\ket{+}=\ket{\text{D}_{5/2},+3/2}$, with amplitude $\sqrt{6/15}$. The $\sigma^+$ ($\sigma^-$) transition is associated with the emission of an R (L) (right and left-hand circular) polarized photon. (HALO = high-numerical-aperture laser objective, PMT = photomultiplier tube, SIA = single-ion-addressing)} \label{fig:setup}
	\end{figure*}

	The schematic of the experimental implementation is shown in \autoref{fig:setup}.
	Two $^{40}\text{Ca}^+$ ions are trapped in a linear Paul trap and the generated photons are collected with two high-numerical aperture laser objectives (HALO NA=$0.4$) along the magnetic field axis. As a result, only the $\sigma^\pm$-polarized components of the emitted photons are collected. One HALO is used to couple the P$_{1/2}$--S$_{1/2}$ fluorescence at 397\,nm into two multi-mode fibers for readout of the state of each atom individually. The second HALO is used to couple the red photons of the P$_{3/2}$--D$_{5/2}$ transition at 854\,nm into two single-mode fibers (780HP). This is accomplished by the use of two telescopes: one 10:1 telescope (telescope 1) mode-matches the emitted photons to the fiber couplers, the second 1:1 telescope (telescope 2) generates an intermediate image of the two ions. The distance of $0.2\,$mm of the two spots is sufficient to separate the two light paths with a D-shaped mirror. More details on the setup are found in appendix \ref{appendix: setup}.
	
	The geometric arrangement of the laser beams is shown in \autoref{fig:setup}(b), and the level-scheme of $^{40}$Ca$^+$ in \autoref{fig:setup}(c). For initial cooling and for state discrimination via fluorescence detection, the 397\,nm and 866\,nm lasers are used. The 729\,nm laser serves for coherent state manipulations on the S$_{1/2}$--D$_{5/2}$ transition, in particular for the two-ion quantum gate. The generation of 854\,nm photons is similar to the procedure described in \cite{APE, Bock2024} and uses a global, $\pi$-polarized 393\,nm laser beam at $90^\circ$ with respect to the magnetic field axis. A 854\,nm laser is used to pump population from D$_{5/2}$ back to the ground state S$_{1/2}$. 
	For this purpose, two beams are installed: one global 854\,nm beam resets both atoms, while the other beam addresses only atom 2 (single-ion-addressing or SIA beam). The addressing beam is coupled in through the same single-mode fiber that collects the 854\,nm photons of atom 2, by means of a 99:1 fiber beam splitter. 
	Additionally a global 866\,nm laser is used to pump population from D$_{3/2}$ back to the ground state S$_{1/2}$, which is necessary due to the parasitic decay channel from P$_{3/2}$ to D$_{3/2}$.

	\subsection{Asynchronous generation of atom-photon entanglement\label{sec:ExperimentATPh}}
	
	The generation of individual atom-photon entanglement is carried out similar to the case with only one atom in \cite{APE, Bock2024}; the contributing levels are shown in \autoref{fig:setup}(d). The process starts by preparing atom $i$ ($i=1,2$) in the ground state S$_{1/2}$ and exciting it with the global $\pi$-polarized 393\,nm laser, triggering 854\,nm-photon emission. 
	Only photon scattering into the Zeeman sublevels  $\ket{+}=\ket{D_{5/2},+3/2}$ and $\ket{-}=\ket{D_{5/2},-1/2}$ is relevant and is filtered by the atomic projection at the end of the protocol. Therefore all other decay channels can be neglected, and photon emission results in the imbalanced entangled state
	\begin{align}
		\ket{\psi}_{i} = \sqrt{\frac{2}{3}} \ket{+}_{i}\ket{\text{L}}_{i}  + \sqrt{\frac{1}{3}}e^{i \omega_L t_{i}} \ket{-}_{i}\ket{\text{R}}_{i} \nonumber \\
	\end{align}
	where $\omega_L = 2\pi \cdot 9.6$\,MHz represents the Larmor frequency between $\ket{+}$ and $\ket{-}$, and $t_{i}$ the time elapsed after emission of the photon. The imbalance of the entanglement due to the different Clebsch-Gordan coefficients is then equalized through a forced population loss in $\ket{+}$ (details in appendix \ref{appendix: protocol}),
	resulting in the maximally entangled state
	\begin{align}
		\ket{\psi}_{i}  &= \sqrt{\frac{1}{2}} \ket{+}_{i} \ket{\text{L}}_{i}   + \sqrt{\frac{1}{2}}e^{i \omega_L t_{i} } \ket{-}_{i} \ket{\text{R}}_{i} \ .\nonumber \\ \label{eq: ape} 
	\end{align}
	Individual atom-photon entanglement generation from atoms 1 and 2 is now asynchronously combined into the sequence of \autoref{fig:seq_protocol}. After initial cooling of the atoms, the sequence starts with the creation of entanglement between atom\,1 and a photon in arm\,A. The global 854\,nm laser beam is used to reset the atom until a photon in arm\,A is detected. This is followed by the generation of entanglement between atom\,2 and a photon in arm\,B. If a trial is not successful, the 854\,nm SIA beam is used to reset only atom\,2, thereby keeping the previously generated state of atom\,1 intact. This cycle is repeated until the second photon is received, or up to $n_\textrm{max}$ times, in which case the protocol aborts and starts from the beginning. 
	
	Before adding the second part of the protocol, we evaluated the individual atom-photon entanglement by full quantum-state tomography. The detection bases of the photons were set by quarter- and half-wave plates and Wollaston prisms in front of four detectors, while atomic state detection used the 729\,nm laser for basis rotations and subsequent fluorescence detection. More details on the individual steps of the protocol are provided in appendix \ref{appendix: protocol}.

	\subsection{Entanglement swapping and generation of photon-photon entanglement}\label{sec:ExperimentPhPh}
	
	To obtain photon-photon entanglement after successful asynchronous generation of two maximally entangled atom-photon states, deterministic projection onto a basis of maximally entangled states is applied to the two atoms. This requires an operation that maps the atoms from the entangled basis $\ket{\Phi^\pm_\text{at,at}}$ and $\ket{\Psi^\pm_\text{at,at}}$ (specified below in \autoref{eq:PPII_basis}) to the measurement basis $\{ \ket{+}\ket{+}, \ket{+}\ket{-}, \ket{-}\ket{+}, \ket{-}\ket{-} \}$. 
	To this end, a quantum gate is implemented between the $\ket{+}$ and $\ket{-}$ states of the two atoms. First a global 729\,nm $\pi$-pulse transfers the $\ket{-}$ populations to $\ket{\text{S}_{1/2},m = +1/2}$. Then a M\o{}lmer-S\o{}rensen gate is applied, acting on the axial sidebands of the $\ket{\text{S}_{1/2},m = +1/2}$ - $\ket{+}$ transition. Finally another global 729\,nm $\pi$-pulse transfers the populations of $\ket{\text{S}_{1/2},m = +1/2}$ back to the $\ket{-}$ state. With this operation (MS), the mapping results as 
	\begin{align} 
		\ket{-}\ket{-} = \text{MS} \left(\ket{-}\ket{-} + i \ket{+}\ket{+} \right) /\sqrt{2} = \text{MS} \ket{\Phi^-_\text{at,at}}\nonumber\\
		\ket{+}\ket{+} = \text{MS} \left(\ket{+}\ket{+} + i \ket{-}\ket{-} \right) /\sqrt{2} = \text{MS} \ket{\Phi^+_\text{at,at}}\nonumber\\
		\ket{+}\ket{-} = \text{MS} \left(\ket{+}\ket{-} - i \ket{-}\ket{+} \right) /\sqrt{2} = \text{MS} \ket{\Psi^-_\text{at,at}}\nonumber\\
		\ket{-}\ket{+} = \text{MS} \left(\ket{-}\ket{+} - i \ket{+}\ket{-} \right) /\sqrt{2} = \text{MS} \ket{\Psi^+_\text{at,at}}\nonumber\\
		\label{eq:PPII_basis}
	\end{align}
	This allows rewriting the (ideal) joint state of the two atom-photon pairs as 
	\begin{align}
		\ket{\psi_\text{joint}} = \ket{\psi}_1 \otimes \ket{\psi}_2 \quad\ \,&\,\nonumber\\ = \frac{1}{\sqrt{8}} \big[+(\ket{\text{L}}\ket{\text{L}} &- e^{i \phi_+} \ket{\text{R}}\ket{\text{R}})\ket{\Phi^+_\text{at,at}}\nonumber\\
		-i\,(\ket{\text{L}}\ket{\text{L}} &+ e^{i \phi_+} \ket{\text{R}}\ket{\text{R}})\ket{\Phi^-_\text{at,at}}\big] \nonumber \\ 
		+\frac{1}{\sqrt{8}} e^{i\phi_\Psi}  \big[
		(\ket{\text{L}}\ket{\text{R}} &+ e^{i \phi_-}\ket{\text{R}}\ket{\text{L}})\ket{\Psi^-_\text{at,at}} \nonumber\\
		+i\,(\ket{\text{L}}\ket{\text{R}} &- e^{i \phi_-} \ket{\text{R}}\ket{\text{L}})\ket{\Psi^+_\text{at,at}}\big] \nonumber\\ \label{eq:PPII_state}
	\end{align} 
	which is transformed by the MS operation to
	\begin{align}
		\text{MS}\ket{\psi_\text{joint}} =
		\frac{i}{\sqrt{8}}\big[+(\ket{\text{L}}\ket{\text{L}} &- e^{i \phi_+} \ket{\text{R}}\ket{\text{R}})\ket{-}\ket{-}\nonumber\\
		-i\,(\ket{\text{L}}\ket{\text{L}} &+   e^{i \phi_+} \ket{\text{R}}\ket{\text{R}})\ket{+}\ket{+}\big]\nonumber \\ 
		+\frac{-i}{\sqrt{8}} e^{i\phi_\Psi}  \big[
		(\ket{\text{L}}\ket{\text{R}} &+ e^{i \phi_-}\ket{\text{R}}\ket{\text{L}})\ket{-}\ket{+}\nonumber\\
		+i\,(\ket{\text{L}}\ket{\text{R}} &-e^{i \phi_-} \ket{\text{R}}\ket{\text{L}})\ket{+}\ket{-}\big].\nonumber\\
		\label{eq: PPII_stateandMS}
	\end{align} 
	The phase values of states (\ref{eq:PPII_state}) and (\ref{eq: PPII_stateandMS}) depend on the times $t_1$ and $t_2$ elapsed after the emission of the two photons and are given by $\phi_-= \omega_L (t_1 - t_2) + \pi/2$, $\phi_+= \omega_L (t_1 + t_2) + \pi/2$. The global phase $\phi_\Psi=\omega_L t_2$ is not visible in the measurements.
	This shows that, depending on the result of the final atom-atom state projection in the $\ket{\pm}$ measurement basis, 
	one out of four maximally entangled photon-photon states is obtained.


	\section{Results}\label{sec:Results}
	
	We present three measurements that characterize our implementation of the QR cell: the first one assesses the individual atom-photon entanglement according to \autoref{sec:ExperimentATPh}. Secondly, the asynchronous generation of photon-photon entanglement according to \autoref{sec:ExperimentPhPh} is evaluated. Finally, the scaling of the protocol with different attenuation levels is described and measured.

	\subsection{Atom-photon entanglement}\label{sec: ExperimentATPhResults}
	
\begin{figure}[h]
		\centering
		\includegraphics[angle=0,scale=0.65]{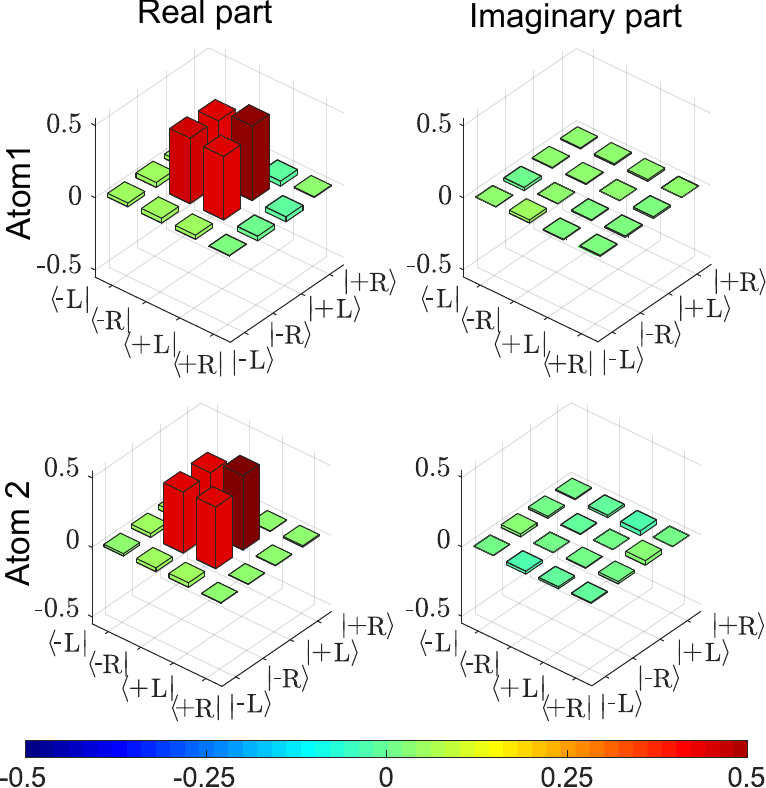}
		\caption{Results of quantum state tomography: density matrices $\rho_1$ and $\rho_2$ of the first and second atom at $n_\textrm{max}$ = 10.}\label{fig_matrices_atom_photon}
		\vspace{\floatsep}
		
		\captionof{table}{Fidelity of the tomographically reconstructed state with the ideal state, $F=\langle\psi_i|\rho_i|\psi_i\rangle$, with $\psi_i$ given by Eq.\,\ref{eq: ape}, and purity $P=\Tr(\rho^2_{i})$ of atom-photon entanglement for atom\,1 and atom\,2 at $n_\textrm{max} = 10$.} 
		\label{tab:APE_fidelity_purity}
		\setlength{\tabcolsep}{.25cm}
		\begin{tabular}{c|c|r} 
			\toprule
			Position  & Fidelity & Purity \\ 
			\toprule
			Atom\,1 & 0.931(5) & 0.88(1)\\ 
			Atom\,2 & 0.924(2) & 0.868(4)\\
			\botrule
		\end{tabular}
	\end{figure}

	Characterization of the two atom-photon entangled states is carried out using quantum state tomography of the individual atom-photon density matrices $\rho_1$ and $\rho_2$ as in \cite{APE}. The experiment is performed with the maximum number of trials for the second atom set to $n_\text{max}=100$, but is evaluated also for lower values of $n_\text{max}$, by post-selecting the events. This allows us to investigate the influence of the waiting time in the asynchronously driven protocol. As an example, the density matrices of the two atom-photon states for $n_\text{max}=10$ are shown in \autoref{fig_matrices_atom_photon}, and the fidelities with the ideal state and their purities are summarized in \autoref{tab:APE_fidelity_purity}. As the time window for photon detection extends over several Larmor precession periods, the Larmor phase is taken into account in the tomography, such that effectively the density matrix for $t=0$ is reconstructed. Accordingly, the finesse is calculated with the state of Eq.\,\ref{eq: ape} for $t=0$.
	
	\begin{figure}[h]
		\centering
		\includegraphics[angle=0,scale=0.6]{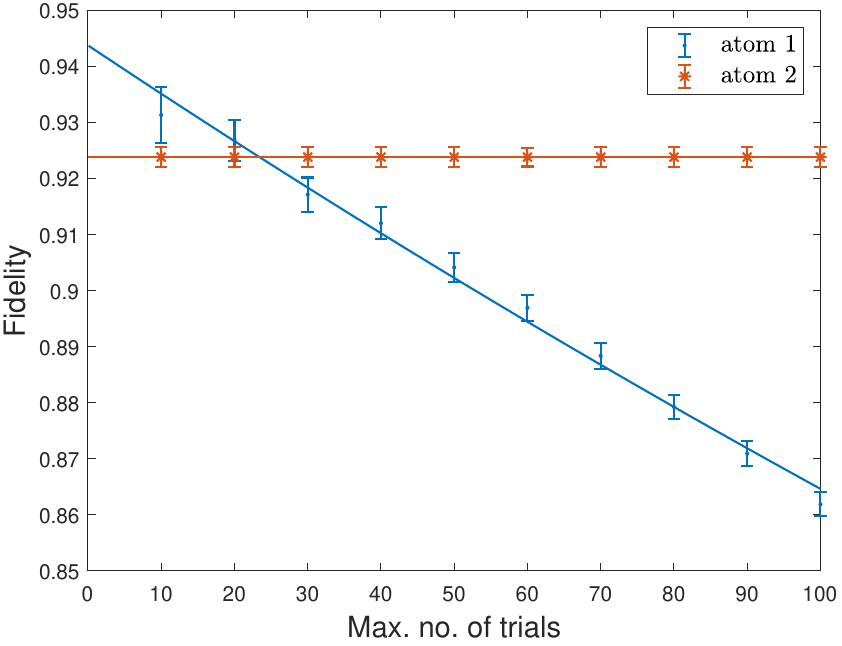}
		\caption{Fidelities of the reconstructed atom-photon density matrices as functions of $n_\textrm{max}$. The lines show a fit for atom\,1, according to the depolarization model, and a constant value for atom\,2. The error bars are generated by a Monte-Carlo method.}\label{fig_fidVsTri_atom_photon}
	\end{figure}

	The fidelities of the two reconstructed atom-photon states depending on the maximum number of trials $n_\textrm{max}$ are plotted in \autoref{fig_fidVsTri_atom_photon}. 
	For atom\,2, a value of $92.4(2)\,\%$ is obtained, shown by the orange points and line. This value is independent of the maximum number of trials, because state tomography is always applied at the same delay after photon\,2 has been detected. 
	For atom\,1 a decline in fidelity is observed. This is understood as a consequence of spurious excitation by the SIA beam (termed false addressing in the following). While this beam resets atom\,2 to S$_{1/2}$ with independently measured probability $P_\textrm{SIA,reset}=99.976(1)\,\%$, also atom\,1 is reset with small probability $P_\textrm{SIA,false} = 0.636(4)\,\%$. The subsequent 393\,nm pulse of the next photon generation cycle causes a mixing of the state of atom\,1 by this amount. In appendix \ref{appendix: ape_ppe_theo}, a model of this mixing process, with the false addressing probability and the initial fidelity as free parameters, is derived. The blue line is a fit to the data with this model, which describes the observation with very good agreement. An initial fidelity of $94.5(2)\,\%$ and a value of $0.56(3)\,\%$ for the false addressing probability are obtained by the fit. 
	The fidelity is slightly lower than what was achieved in a previous experiment \cite{APE}; this is attributed to a slight miscalibration of the magnetic field axis with respect to the direction of photon collection. Moreover, the fidelity value for atom\,2 is lower than that of atom\,1, which may be caused by imperfections in the optical elements and the polarization projection setup of arm\,B.
	The independently measured value for the false addressing $P_\textrm{SIA,false}$ differs by two to three standard deviations from the fitted value of 0.56(3)\,\%, which indicates that some systematic discrepancies have not been accounted for in the statistics. A possible explanation are environmental changes during the measurements.
	
	We use \autoref{fig_fidVsTri_atom_photon} to estimate the maximum number of trials for which a fidelity of 50\,\% is still obtained. From the fitted dependency this absolute maximum is 1442 trials.

	\subsection{Photon-photon entanglement} \label{sec: ExperimentPhPhResults}
	
	\begin{figure}[h]
		\centering
		\includegraphics[angle=0,scale=0.56]{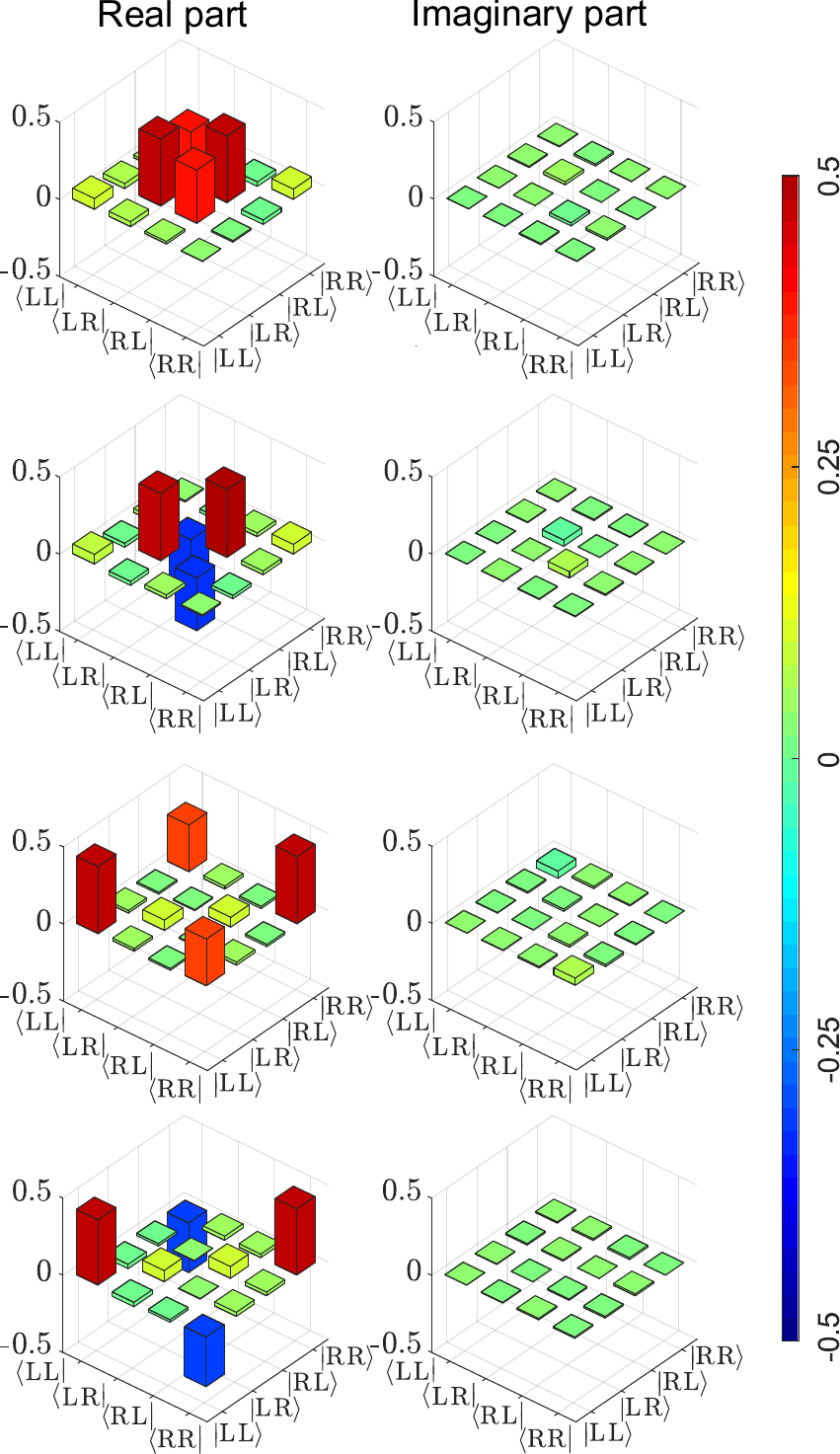}
		\caption{Tomographically reconstructed density matrices of the photon-photon states for the four projection results of the joint atom-atom state ($\rho_{meas}$) corresponding to $\ket{\Psi^-_\text{at,at}}, \ket{\Psi^+_\text{at,at}}, \ket{\Phi^-_\text{at,at}}$ and $\ket{\Phi^+_\text{at,at}}$ (top to bottom) with $n_\text{max}$ = 10. 
		}\label{fig_matrices}
		\vspace{\floatsep}
		\centering
		\captionof{table}{Fidelity and purity of photon-photon states for the four projection results of the atom-atom state, 
			at $n_\text{max}$ = 10.}
		\label{tab:PPE_fidelity_purity}
		\setlength{\tabcolsep}{.25cm}
		\begin{tabular}{c|c|r} 
			\toprule
			Projected state & Fidelity & Purity \\ 
			\toprule
			$\ket{\Psi^-_\text{at,at}}$ & 0.782(4) & 0.632(7)\\ 
			$\ket{\Psi^+_\text{at,at}}$ & 0.778(6) & 0.627(7)\\
			$\ket{\Phi^-_\text{at,at}}$ & 0.739(4) & 0.576(7)\\ 
			$\ket{\Phi^+_\text{at,at}}$ & 0.758(6) & 0.588(7)\\
			\botrule
		\end{tabular}
	\end{figure}
	The previously established and characterized atom-photon entanglement is now utilized in the swapping procedure of \autoref{sec:ExperimentPhPh} to generate photon-photon entanglement. Subsequent quantum state tomography on the two photons is carried out to reconstruct their density matrix. 
	
	As in the previous section, the experiment is performed with the maximum number of trials for the second atom set to $n_\textrm{max}=100$, and evaluated for lower values of $n_\textrm{max}$ by post-selecting the events, in order to infer the influence of the asynchronous protocol. The resulting density matrices for $n_\textrm{max}=10$ are displayed in \autoref{fig_matrices}, separately for the four results of the atom-atom projection corresponding to $\ket{\Psi^\pm_\text{at,at}}$ and $\ket{\Phi^\pm_\text{at,at}}$ (see Eqs.~(\ref{eq:PPII_basis}) and (\ref{eq: PPII_stateandMS})). The corresponding values of fidelity and purity are listed in \autoref{tab:PPE_fidelity_purity}.
	
\begin{figure}[h]
		\centering
		\includegraphics[scale=0.59]{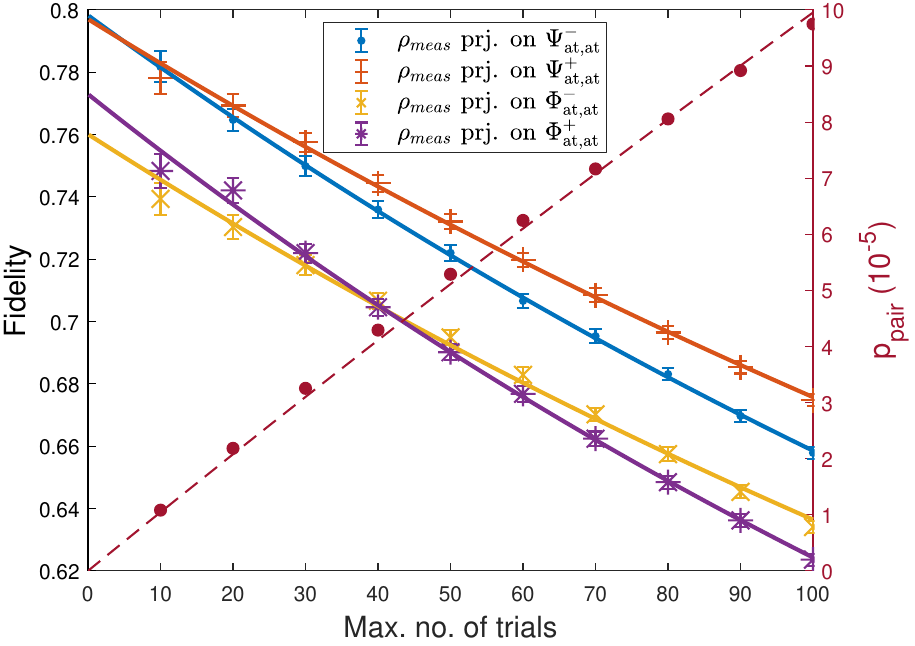}
		\caption{Fidelity values, calculated from the tomographically reconstructed photon-photon states, depending on the maximum number of trials $n_\textrm{max}$, for the different outcomes of the atom-atom state projection. Solid lines are fitted model calculations, explained in the text. The dark red dots are the measured photon pair detection probabilities, $p_\text{pair}$, the dashed line is a guide to the eye.}\label{fig_FidelityVsTrials}
		\captionof{table}{Fidelities of the M\o{}lmer-S\o{}rensen gate, of the probability of false addressing, and initial fidelity, as derived from the fits in Fig.~\ref{fig_FidelityVsTrials}.} 
		\setlength{\tabcolsep}{.25cm}
		\label{tab:PPE_fidelity_fitresult}
		\begin{tabular}{c|c|c|r} 
			\toprule
			Projected state & $\mathcal{F}_\textrm{MS}$ & $P_\textrm{SIA,false}$ & $\mathcal{F}_\textrm{init}$\\ 
			\toprule
			$\ket{\Psi^-_\text{at,at}}$ & 0.915(1) & 0.0130(2) & 0.797(1)\\ 
			$\ket{\Psi^+_\text{at,at}}$ & 0.914(2) & 0.0110(2) & 0.796(2)\\
			$\ket{\Phi^-_\text{at,at}}$ & 0.871(3) & 0.0120(4) & 0.759(3)\\ 
			$\ket{\Phi^+_\text{at,at}}$ & 0.886(3) & 0.0146(4) & 0.771(3)\\
			\botrule
		\end{tabular}
	\end{figure}

	\autoref{fig_FidelityVsTrials} shows the decline of the photon-photon entangled-state fidelity with increasing $n_\textrm{max}$; for comparison, the probability of photon pair detection, $p_\text{pair}$, is also plotted. 
	The model of the mixing process of appendix \ref{appendix: ape_ppe_theo} is used to fit the data (solid lines), taking into account an imperfect M\o{}lmer-S\o{}rensen gate with a certain average state fidelity $\mathcal{F}_\textrm{MS}$. The model uses the initial fidelities of the two atom-photon states of \autoref{sec: ExperimentATPhResults} as input, and the M\o{}lmer-S\o{}rensen gate fidelity $\mathcal{F}_\textrm{MS}$ and the false addressing probability $P_\textrm{SIA,false}$ as free parameters. The results of the four fits are summarized in \autoref{tab:PPE_fidelity_fitresult}. The initial fidelity $\mathcal{F}_\textrm{init}$, corresponding to the fidelity with only one trial, is then calculated from the fitted model. 
	
	We can compare the results of \autoref{tab:PPE_fidelity_fitresult} with values for the same quantities that were independently measured in the course of this experimental run: the M\o{}lmer-S\o{}rensen gate fidelity was found to be $\mathcal{F}_\text{MS}=92.6(17)\,\%$ (see appendix \ref{appendix: MSgate}), and the false addressing and reset probabilities were $P_\text{SIA,false} = 0.9(3)\,\%$ and $P_\text{SIA,reset}>99\,\%$. 
	One notices that the M\o{}lmer-S\o{}rensen gate fidelity and the false addressing probability are different for the four projection results, and slightly different from the independently measured values. We attribute this mainly to decoherence induced by magnetic field fluctuations, which are not included in the model. One understands their effect considering the time dependence of the phase values of the states given in \autoref{sec:ExperimentPhPh}. For the $\ket{\Psi^\pm_\text{at,at}}$ states the difference of the photon detection times enters, causing a cancellation of the effect of the magnetic field fluctuations in the period between the detection of photon\,2 and the MS operation. For the $\ket{\Phi^\pm_\text{at,at}}$ states, the sum of the detection times enters and the effect of the fluctuations during this period is amplified.
	
	The model allows us to infer the maximum number of trials before the 50\,\% fidelity threshold \cite{Sackett_2000} is reached. We find it to be 358 in the best case of $\ket{\Psi^+_\text{at,at}}$, and 240 in the worst case of $\ket{\Phi^+_\text{at,at}}$. As expected, this number is lower than the 1442 possible trials for double atom-photon entanglement. The main reasons are the slightly larger false addressing probability during this experiment, the additional M\o{}lmer-S\o{}rensen gate with fidelity less than 1, and that the individual infidelities of the separate atom-photon states now combine into an overall initial infidelity before the M\o{}lmer-S\o{}rensen gate.  

	\subsection{Scaling}\label{sec: ExperimentScalingResults}
	
	
	In this section, we investigate the scaling of our QR cell implementation, i.e., the success probability for generating photon-photon entanglement, as well as its rate, depending on the transmission of the photonic channel. A real communication scenario would also require additional communication time in the protocol, i.e., the time the photon travels to the receiver and the time needed to inform the central station of the detection signal. This communication time is not considered at this point.
	
	To simulate channel attenuation, we place optical density filters in the two photon paths in front of the detectors. Four examples are realized: transmission of 100\% (no filter), 78\%, 48\% and 24\%.
	
\begin{figure}[h]
		\centering
		\includegraphics[scale=0.6]{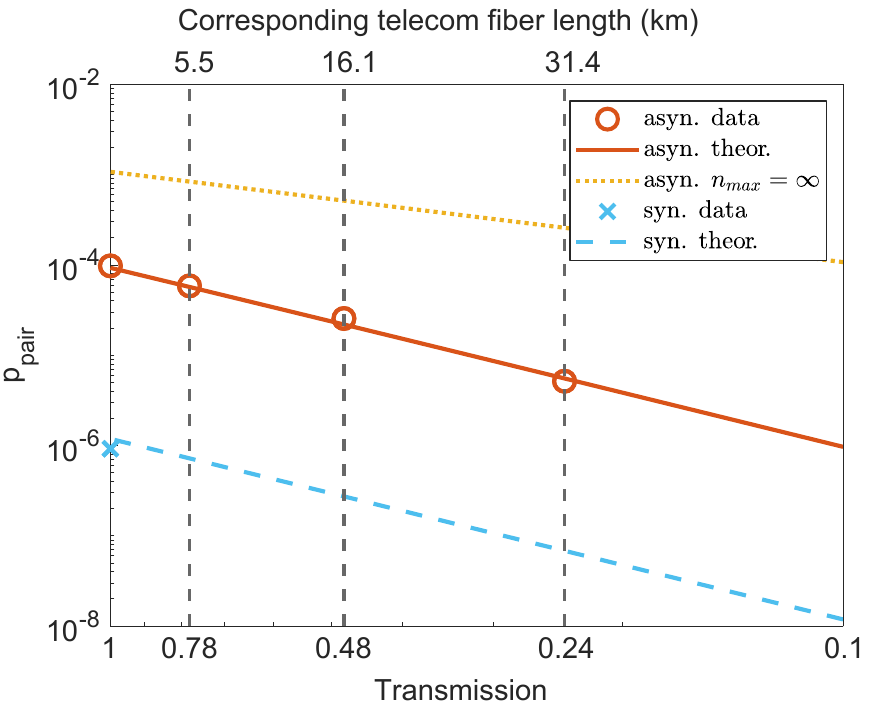}
		\caption{Photon pair detection probability for different channel transmission (bottom axis) and corresponding telecom fiber length (top axis). Shown are the measured probabilities for the asynchronous case with $n_\text{max}=100$ (red rings) and the calculated curve (red solid line) with independently measured $p_1$ and $p_2$ (Eq.~(\ref{eq:asny})). The measured synchronous probability with no attenuation (blue cross) is plotted together with the theoretical synchronous behavior according to Eq.\,(\ref{eq:sny}) (blue dashed line). The ideal behavior at $n_\text{max}\rightarrow \infty$ is also plotted (yellow dotted line), marking the theoretical limit.} \label{fig_scaling}
		\vspace{\floatsep}
		
		\captionof{table}{Measured probability (with $n_\text{max}=100$) to detect a photon-photon pair per repetition of the protocol depending on the transmission, and measured rates of detected photon-pairs per second, without considering additional communication time of the asynchronous protocol added by the corresponding values for the synchronous protocol.}
		\label{tab1}
		\setlength{\tabcolsep}{.25cm}
		\begin{tabular}{l|c|r} 
			\toprule
			Transmission & $p_{pair}$ & Rate (s$^{-1}$) \\
			\toprule
			asyn. 24\% & $5.15(3)\cdot 10^{-6}$ & 0.77(1)\\ 
			asyn. 48\% & $2.55(1)\cdot 10^{-5}$ & 3.44(1)\\
			asyn. 78\% & $5.86(1)\cdot 10^{-5}$ & 7.46(2)\\ 
			asyn. 100\% & $9.76(2)\cdot 10^{-5}$ & 11.34(2)\\ 
			\toprule
			syn. 100\% & $9.2(3)\cdot 10^{-7}$ & 0.230(6)\\ 
			\botrule
		\end{tabular}
	\end{figure}

	First we examine the measured probabilities for detecting a photon-photon pair, $p_\text{pair}$. They are shown in \autoref{fig_scaling} (red rings) and listed in \autoref{tab1} for a maximum number of $n_\text{max}=100$ trials. For comparison, the photon pair probability is also measured for synchronous generation (blue cross), which corresponds to the case $n_\text{max} = 1$. The telecom fiber length corresponding to the inserted filter transmission is calculated for 1550\,nm fiber with 0.2\,dB/km attenuation. Conversion of our 854\,nm photons to this wavelength has been achieved by quantum frequency conversion with up to $\sim 60$\,\% efficiency \cite{Elena1, Bock2024, Kucera2024}. Taking this conversion efficiency into account reduces the corresponding telecom fiber length by 11.1\,km, but this not considered in the general case plotted in \autoref{fig_scaling}. 
	
	In one repetition of the asynchronous protocol (defined as excitation until double photon detection, or until $n_\text{max}$ is reached), the probability for detecting a photon pair is given by 
	\begin{align}
		p_\text{pair,asyn}=p_1(1-(1-p_2)^{n_\text{max}}) \label{eq:asny}
	\end{align}
	with $p_1$ ($p_2$) denoting the probability of generating and detecting the first (second) photon for a single 393\,nm excitation pulse. 
	In the case of the synchronous protocol
	the probability is given by 
	\begin{align}
		p_\text{pair,syn} = p_1 \cdot p_2 \label{eq:sny}
	\end{align}
	which results from Eq.\,(\ref{eq:asny}) for $n_\text{max} = 1$. For the theoretical curves in \autoref{fig_scaling}, the probabilities $p_{1} = 0.114\,\%$ and $p_{2} = 0.096\,\%$ are taken from an independent measurement (details in appendix \ref{appendix: telescope}). The three lines show the scaling for our measurement, $n_\text{max} = 100$, for the synchronous case, $n_\text{max} = 1$, and for the theoretical upper limit of the asynchronous protocol, $n_\text{max}\rightarrow \infty$, when the second photon is always detected. In our case, the asynchronous protocol shows a 100-fold improvement over the synchronous case. A reduced slope of the line, however, which would be an important aspect of the repeater advantage, is not yet observable. It would require a significantly larger number of trials, as discussed further below.
	
	
	The second aspect is the behavior of the rates of photon-photon pairs, calculated from the measured detection events and the measurement time. The measurement time includes the experimental overhead consisting of the cooling time and the time for projection pulses, if an attempt was successful. The experimentally obtained rates are summarized in \autoref{tab1}. The rates do not scale like the detection probabilities of the pairs, as a consequence of the included overhead. 
	The comparison of the rates for the asynchronous and synchronous cases at 100\% transmission, 11.34(2)\,s$^{-1}$ and 0.230(6)\,s$^{-1}$ respectively, shows a 49-fold increase of the rate by using the asynchronous protocol.

	\section{Summary and Discussion}
	
	In this work, the implementation of an asynchronously operated quantum repeater cell, based on single photons from two single trapped ions, is demonstrated and characterized. A 100-fold improvement of the photon pair probability compared to a synchronously operated QR cell highlights the advantage of this protocol. With the maximum number of asynchronous trials set to $n_\text{max}=10$, a photon-photon entanglement fidelity of $77.8(6)\,\%$ is measured, while with $n_\text{max}=100$ the fidelity is still $67.5(2)\,\%$. We infer that for $n_\text{max} \leq 358$ asynchronous trials, the entanglement fidelity will remain above $50\,\%$. 
	
	
	
	In the following, we will assess how close our implementation is to demonstrating a genuine quantum repeater advantage. For that, we compare the rate and fidelity of this asynchronous implementation, where two atoms send their photons over half the communication distance, with the 
	alternative that one single atom transmits its entangled photon \cite{APE} across the whole distance. We consider a realistic case of $2\times31.4\,$km telecom fiber (24\,\% fiber transmission per half distance). 
	%
	%
	The general formalism of \cite{Luetkenhausprotokoll} is adapted to this implementation and a rate estimation is derived in appendix \ref{appendix:rate_model}. When the efficiency of photon generation per trial is assumed equal for the two setups, we find that a rate advantage of the asynchronous protocol requires $n_\text{max}\geq 1380$. Since this implies that for the presented setup the entanglement fidelity falls below $50\,\%$, it will be necessary to either increase the photon generation and/or collection probability per attempt or to improve the fidelity, or both. 
	
	A first obvious step is using a single-ion addressing beam also for atom\,1, which effectively doubles the pair rate, and thereby decreases the threshold for a rate advantage to $n_\text{max} \geq 573$ according to the rate estimations; but this still leads to a fidelity below $50\,\%$. Additional replacement of the current HALO by an $\text{NA}=0.7$ objective would result in a 3.4 times higher collection probability, yielding a rate advantage at $n_\text{max}\geq 169$ which is within reach with this experiment. 
	
	
	Improvement of the fidelities of the individual operations, i.e., of the M\o{}lmer-S\o{}rensen gate and the individual atom-photon entanglement, would offset the curves of \autoref{fig_FidelityVsTrials} towards unity fidelity at the beginning. Such improvement is possible and demonstrated in similar experiments \cite{Gaebler2016, APE}. But we do not expect a large impact, as these values are already well above 90\,\%.
	
	The most severely fidelity-limiting factor of the present setup is false addressing, i.e., spurious excitation of the wrong atom in the single-ion addressing. Although a Gaussian beam with 0.8\,$\mu$m waist (1/e field radius) at the atom's position theoretically results in negligible false addressing, the laboratory observation is that residual aberrations of the optical elements prevent us from reaching this limit. Unfortunately, no practical improvement by a redesign of the optical system is in reach. 
	Increasing the distance between the ions is not a solution either; we found that in the regime where the axial sideband frequencies ensure good M\o{}lmer-S\o{}rensen gate performance, our current beam profile does not permit to diminish the crosstalk significantly.
	Nevertheless, by using a 0.7-NA objective and maximizing the fidelity of atom-photon entanglement and M\o{}lmer-S\o{}rensen gate, the average fidelity would saturate at 53.6\,\%, meaning that the photon pair will always retain some entanglement, because the pair generation probability converges to 1 before the fidelity drops below 50\,\%.
	
	An approach that offers serious improvement of the photon collection and generation efficiency is using a cavity. It enabled achieving a repeater advantage in the two recent realizations of a QR cell with neutral atoms \cite{PhysRevLett_Rempe} and with ions \cite{Ion_Innsbruck}. The latter implementation uses a 20\,mm-long cavity and demonstrated a rate of 5.9\,s$^{-1}$ photon pairs with 72.2(2)\,\% entanglement fidelity after 25\,km of fiber ($\sim40$\,\% transmission) in both photon arms, and including the communication time of $\sim250\,\mu$s. 
	While clearly advantageous to our implementation without cavity, there is also a caveat to the long cavity in \cite{Ion_Innsbruck}, in that its $1.14\,\mu$s ringdown time poses a limit on how efficient a channel capacity may be used. For a single device this makes no difference, as the communication time is still much larger for reasonable distances of a repeater. But in the perspective of using copies of such a device or implementing more complex repeater schemes involving multiple memory qubits in the same trap, this timescale enters and a shorter cavity \cite{Teller_2023, Steiner_2013, Takahashi_2013} will be desirable. Besides the collection efficiency and the achievable generation rate, use of a short cavity will also improve the photon purity, which is relevant for photonic Bell-state measurement on a beam splitter. A sub-mm cavity integrated with a linear ion trap is under construction in our group, in order to approach this next step in the development of quantum repeater technology. 
	
	
	
	
	\begin{acknowledgments}
		We acknowledge support by the German Federal Ministry of Education and Research (BMBF) through projects Q.Link.X (16KIS0864) and QR.X (16KISQ001K).
		
		M.B. and O.E. contributed equally to this work. M.B., O.E., S.K. and M.K. set up the experiments; M.B. and O.E. performed the experiments and analyzed the data with evaluation scripts written by S.K. and M.K.; M.B., O.E. and S.K. wrote the paper with input from all authors; J.E. and S.K. conceived and supervised the project.
	\end{acknowledgments}

	\appendix
	\section{Details of the experimental setup}\label{appendix: setup}
	\subsection{Individual atom readout}
	To detect the 397\,nm fluorescence of both atoms individually, the two atoms are imaged onto two bare multimode fibers which are mount in a single ferrule (blue path in \autoref{fig:setup}). The photons are counted using photo-multiplier tubes. The cross talk between the two individually coupled beams is measured by trapping a single atom at the position of atom\,1 (atom\,2) and counting the fiber coupled photons of both fibers. The cross talk is then the ratio of the counts from the two position for each fiber. This results for the atom\,1 (atom\,2) fiber $P_\textrm{false,at1} = \frac{200\,\text{cts/s}}{110\,\text{kcts/s}} = 1.8\times 10^{-3}$  ($P_\textrm{false,at2} = \frac{400\,\text{cts/s}}{150\,\text{kcts/s}} = 2.7\times 10^{-3}$).
	
	\subsection{854\,nm telescopes and beam separation}\label{appendix: telescope}
	To separately couple the 854\,nm photons into two single-mode fibers, first a 1:10 telescope ($f=300\,$mm \& $f=30\,$mm) and second a 1:1 telescope ($f=100\,$mm) is used. Starting from the measured axial sideband frequency of $\nu_\text{ax}=1.1642(2)$\,MHz, a distance of the two atoms of 5.1\,$\mu$m is calculated \cite{AbstandIonen}. With the focal length of the objective ($25\,$mm), a separation of the two atoms at the image in the center of the 1:1 telescope of $0.2\,$mm is sufficient to separate the two beams by a D-shaped mirror. The cross talk between the two individually coupled beams is measured by trapping a single atom at the position of atom\,1 and counting the fiber coupled photons of both fibers in a pulsed way. The procedure is repeated with the atom at the position of atom\,2. As a result, the cross talk of the two atoms is bound by $<6\cdot 10^{-6}$.
	
	The fiber coupling efficiency is independently measured by excitation with a $\pi$-polarized 393\,nm beam and collection of the fiber coupled 854\,nm and 850\,nm photons. The ratio of collected photons and excitation trials is measured to $0.63\,\%$ ($0.58\,\%$) for atom\,1 (atom\,2) by bypassing the polarization projection setup. The branching ratio $\eta_{850} = \frac{\text{A}_{854}}{\text{A}_{850}+\text{A}_{854}} = 89.9\,\%$ corrects for the collected 850\,nm photons ($\text{A}_{854} = 1.35$\,MHz and $\text{A}_{850} = 0.152$\,MHz). $\eta_{mix} = 50\,\%$ due to initially starting in a mixed state, $\eta_{sigma}=9/15$ is the fraction of $\sigma$-polarized 854\,nm photons, which are collected by the objective with $\eta_{HALO}=6\,\%$, and $\eta_{balance} = 2/3$ because of the treatment of the imbalance due to the different Clebsch-Gordan coefficients. In the experiment two different gate-windows are used for the detection of the photons of atom\,1 (atom\,2) $\eta_{gate} \approx 100\,\%$ ($82\,\%$).
	By applying the above listed correction factors, the fiber coupling efficiency is determined to be $\eta_{fiber} = 19.3\,\%$ ($17.7\,\%$).
	
	In total this leads to a detection efficiency for atom\,1 (atom\,2) of $0.114\,\%$ ($0.096\,\%$). The contributions to this efficiency are summaized in \autoref{tab:detection_eff}.
	
	\begin{table}[h]
		\centering
		\setlength{\tabcolsep}{.25cm}
		\caption{Contributions to the detection efficiency of photons emitted by the two atoms.}
		\label{tab:detection_eff}
		\begin{tabular}{l|r|r}
			\toprule
			Contribution to det. efficiency & Atom 1 & Atom 2\\
			\toprule
			$\eta_{850}$ & 89.9\,\% & 89.9\,\%\\
			$\eta_{mix}$  & 50\,\% & 50\,\%\\
			$\eta_{sigma}$ & 9/15& 9/15\\
			$\eta_{HALO}$ & 6\,\% & 6\,\% \\
			$\eta_{balance}$ & 2/3 & 2/3\\
			$\eta_{gate}$ & 100\% & 82\%\\
			$\eta_{fiber}$ & 19.3\% & 17.7\% \\
			Transm. of projection setup & 60.3\% & 67.2\%\\ 
			Detector efficiency & 91\% & 91\%\\
			\botrule
			$\Pi$ &  0.114\% & 0.096\% \\
			\botrule
		\end{tabular}
	\end{table}

	\subsection{SIA-beam switching}
	The SIA-beam is controlled by two consecutive acousto-optic modulators (AOM) to provide a high extinction ratio for the switching. The necessity of the high extinction has two reasons, firstly that the atom is not excited and repumped by accident, secondly a fraction of the laser light is guided to the detector due to reflections at optical elements and produces background during the time of detection window. By measuring the necessary power for the reset pulse at a given pulse duration, and the back-coupled fraction, the required extinction ratio of $10^{-10}$ is determined. The setup consists of a free space AOM where the first diffraction order is coupled into a fiber, and a fiber-AOM. The measured  extinction ratio of $r_\text{free}=2.76\cdot10^{-6}$ and $r_\text{fiber}=1.29\cdot10^{-7}$ result in a total system extinction ratio of $r_\text{total}= 1.27 \cdot10^{-12}$.

	
	\section{Details on the implementation of the protocol}\label{appendix: protocol}
	The following list contains details on the sequential implementation of the protocol:
	\begin{itemize}
		\item[(1)] $[ duration : 3\,\mu\text{s}]$ Doppler cooling with 397\,nm, 866\,nm, in combination with repumping to S$_{1/2}$ with global 854\,nm. (next$\rightarrow$(2))
		
		\item[(2)] $[ duration : 2\,\mu\text{s}]$ 854\,nm photon generation of both atoms with the $1.7\,\mu\text{s}$ $\pi$-polarized 393\,nm laser pulse and $0.3\,\mu\text{s}$ waiting for detection. The sequence jumps to (3) conditioned on a detection of a photon emitted by atom\,1, otherwise to (1).
		
		\item[(3)] $[ duration : 50\,\mu\text{s}]$ Fluorescence detection to detect the decay of atom\,1 to D$_{3/2}$ at 850\,nm which cannot be distinguished by an 854\,nm decay by the detection setup. Therefore, the lasers (397\,nm and 866\,nm) are switched on for fluorescence detection and the detected 397\,nm photons are counted. A bright detection indicates the population in D$_{3/2}$ and the sequence jumps to (1) otherwise the sequence continues with step (4).
		
		\item[(4)] $[ duration : 4.5\,\mu\text{s}]$ Repreparation of atom\,2 in S${_{1/2}}$ by the SIA-beam and the 866\,nm laser. (next$\rightarrow$(5))
		
		\item[(5)] $[ duration : 2\,\mu\text{s}]$ 854\,nm photon generation of atom\,2 with the $\pi$-polarized 393\,nm laser. The sequence jumps to (6) conditioned on a detection of a photon emitted by atom\,2, if the maximum number of trials $n_\text{max}$ is reached to (1), otherwise to (4) for the next trial.
		
		\item[(6)] $[ duration : 10\,\mu\text{s}]$ Pumping the S$_{1/2}$ population of both atoms to D$_{3/2}$ with the 397\,nm laser. This step, in combination with a dark result of the fluorescence detection of step (10) causes a projection of the population onto D$_{5/2}$. This removes mainly dark-count events in the statistics. (next$\rightarrow$(7))
		
		\item[(7)] $[ duration : 10\,\mu\text{s}]$ A  729\,nm $\pi/2$-pulse transfers  50\,\% of the $\ket{+}$ population of both atoms to S$_{1/2}$ to compensate the imbalance caused by the Clebsch-Gordan coefficients. (next$\rightarrow$(8))
		
		\item[(8)] $[ duration : 10\,\mu\text{s}]$ Pumping the S$_{1/2}$ population of both atoms to D$_{3/2}$ with the 397\,nm laser. This step, in combination with the fluorescence detection of step (10) causes a projection of the population onto D$_{5/2}$ which finalizes the treatment of the imbalance. (next$\rightarrow$(9))
		
		\item[(9)] $[ duration : 220\,\mu\text{s}]$ M\o{}lmer-S\o{}rensen gate procedure:
		A 729\,nm $\pi$-pulse transfers the $\ket{-}$ population to $\ket{\text{S}_{1/2},m = +1/2}$ $[10\,\mu\text{s}]$. The M\o{}lmer-S\o{}rensen gate is applied on the axial sidebands of the $\ket{\text{S}_{1/2},m = +1/2}-\ket{+}$ transition $[200\,\mu\text{s}]$. A 729\,nm $\pi$-pulse transfers the $\ket{\text{S}_{1/2},m = +1/2}$ population to $\ket{-}$ $[10\,\mu\text{s}]$. (next$\rightarrow$(10))
		
		\item[(10)] $[ duration : 100\,\mu\text{s}]$
		Fluorescence detection to detect the population in  D$_{3/2}$. Therefore, the lasers (397\,nm and 866\,nm) are switched on for fluorescence detection and the detected 397\,nm photons are counted. A bright detection indicates the population in D$_{3/2}$ and the sequence is aborted and jumps to (1), otherwise the sequence continues with step (11).
		
		\item[(11)] $[ duration : 110\,\mu\text{s}]$ A 729\,nm $\pi$-pulse transfers the $\ket{-}$ population of both atoms to S$_{1/2}$ $[10\,\mu\text{s}]$. A bright result in a following fluorescence detection $[100\,\mu\text{s}]$ projects onto $\ket{-}$ and the sequence is finished, otherwise the sequence continues with step (12).
		
		\item[(12)] $[ duration : 110\,\mu\text{s}]$ A 729\,nm $\pi$-pulse transfers the $\ket{+}$ population of both atoms to S$_{1/2}$ $[10\,\mu\text{s}]$. A bright result in a following fluorescence detection $[100\,\mu\text{s}]$ projects onto $\ket{+}$ and the sequence is finished.
	\end{itemize}
	The projections of steps (11) and (12) onto the $\ket{\pm}$ states of both atoms allow the distinction of the relevant decay channels for the evaluation.

	\section{Influence of false addressing on fidelity}\label{appendix: ape_ppe_theo}
	
	A theoretical model to predict the decline of fidelity of the atom-photon and photon-photon entanglement caused by false addressing by the SIA-beam is derived. This model is used to fit the experimental results from the state reconstruction and uses the following abbreviations:
	\begin{itemize}
		\item $\rho_{1/2,0}$ : initial atom-photon state\,1 \& \,2 density matrices.
		\item $N = n_\text{max}$ : maximum number of trials.
		\item $p$ : single shot detection probability of a single photon. 
	\end{itemize}
	To address atom\,2, the SIA-beam and the 866\,nm laser is switched on. The false addressing then causes population transfer of D$_{5/2}$ to S$_{1/2}$ of atom\,1, and the sequential excitation with the global 393\,nm causes a mixing of the state of atom\,1 as a consequence. This $\pi$-polarized beam distributes the population to the four Zeeman substates of D$_{5/2}$ ($\ket{D_{5/2},-3/2}$, $\ket{D_{5/2},-1/2} =\ket{-}$, $\ket{D_{5/2},+1/2}$, and $\ket{D_{5/2},+3/2} =\ket{+}$) and due to the 850\,nm decay also to D$_{3/2}$ which is accounted by the branching ratio to D$_{5/2}$ of $\eta_{850} = 89.9\,\%$ (see appendix \ref{appendix: telescope}). The probability to transfer the population to $\ket{\pm}$ is therefore $c = (P_\text{SIA,false}\, \eta_{850})/2$. The distribution of this population due to the Clebsch-Gordan coefficients leads to the depolarizing matrix
	\begin{equation}
		M = \left(\frac{3}{5} \ket{-}\bra{-} + \frac{2}{5} \ket{+}\bra{+}\right) \otimes \frac{1}{2}\mathbb{I}_{photon}
	\end{equation}
	whereby also the $\pi$-decay $\ket{P_{3/2},-1/2}$ to $\ket{-}$ is taken into account. It should be noted that this distribution has no effect onto the calculated fidelity.
	The effect of one trial onto the $i$-th density matrix of atom-photon state\,1 ($\rho_{1,i}$) is then described by the recursive depolarization
	\begin{equation}
		\varepsilon(\rho_{1,i}) = (1-c) \rho_{1,i} + c M\ .
	\end{equation}
	Applying this depolarization $k$ times onto the initial $\rho_{1,0}$ density matrix accounts for all photon generation trials, and leads to the density matrix
	\begin{align}
		\varepsilon_k(\rho_{1,0}) = (1-c)^{k} \rho_{1,0} + (1-(1-c)^{k}) \,M\ .\label{eq: density_atom1_ape}
	\end{align}
	This leads to the fidelity with the ideal state $\ket{\Psi}$ of the output state after $k$ photon generation trials 
	\begin{align}
		\label{eq: fidelity_model_ape}
		\mathcal{F}_1 (k) = (1-c)^{k}\,\mathcal{F}_{1,0} + (1-(1 - c)^{k})\,\frac{1}{4}
	\end{align}
	where the initial fidelity is given by $\mathcal{F}_{1,0} = \bra{\Psi}\rho_{1,0}\ket{\Psi}$, and $\bra{\Psi}M\ket{\Psi} = \frac{1}{4}$ is used. It should be noted that in the protocol of \autoref{sec:ExperimentPhPh}, the procedure to balance the state \ref{eq: density_atom1_ape} is applied, which has no influence on the derived model. 
	
	The experimental protocol for the asynchronous atom-photon entanglement generation is finished once the second photon (i.e. the photon of atom\,2) is detected. To account for the maximum number of trials $N$, an average fidelity is calculated with the weighted average 
	\begin{equation}\label{eq: weighted average fidelity}
		\mathcal{F}_{1}(N) = \sum_{k=1}^{N} \mathcal{F}_1(k) \,w_k\ .
	\end{equation}
	The weights $w_k$ account for the probability to detect the second photon in the $k$-th trial. This is given by 
	\begin{equation}
		w_k = \frac{p\,(1-p)^{k-1}}{\sum_{i=1}^{N} p\,(1-p)^{i-1}} .
	\end{equation}
	The modeled fidelity for the atom-photon entanglement of atom\,1  (\autoref{sec: ExperimentATPhResults}) is then 
	\begin{align}
		\label{eq:weighted average ape}
		&\mathcal{F}_{1}(N) = \nonumber\\
		&\ \frac{1}{4} + \left(\mathcal{F}_{1,0}-\frac{1}{4}\right) \frac{p(1-c)}{1-(1-p)^{N}}\ \frac{1-(1-c)^N(1-p)^N}{p+c\left(1 - p\right)}\ .
	\end{align}
	
	The fidelity $\mathcal{F}_{MS}$ of the applied M\o{}lmer-S\o{}rensen gate ($\text{MS}(\ldots)_{real}$) is included into the model by treating the errors as depolarizing channel
	\begin{align}
		\text{MS}(\rho_{1}\otimes \rho_{2})_{real} = &\left(1-\alpha\right)\text{MS}(\rho_{1}\otimes \rho_{2})_{ideal}\nonumber \\
		&\ \  +\frac{\alpha}{16} \left(\mathbb{I}_{1}\otimes\mathbb{I}_{2}\right) \\ 
		\text{with:}\ \ \ \alpha = &  \frac{16}{15}\left(1-\mathcal{F}_{MS}\right) \ . \nonumber
	\end{align}
	whereby $\rho_{1/2}$ denote the density matrices of the two atom-photon states, and $\text{MS}(\ldots)_{ideal}$ the ideal performing gate. The atom-photon density matrices are also treated in the sense of depolarization of the ideal states $\ket{\Psi}_{1/2}$ of \autoref{eq: ape}
	\begin{align}
		\rho_{i,0} = &(1-q_i) \ket{\psi}\bra{\psi}_i  +\frac{q_i}{4}\mathbb{I}_{i} \\ 
		\text{with:}\ \ \ q_i = &  \frac{4}{3}\left(1-\mathcal{F}_{i,0}\right) \ . \nonumber
	\end{align}
	The averaged fidelity with an ideal state $\ket{\Phi}$ of the two atom-photon states after the full application of the protocol is then calculated by
	\begin{align}
		\label{eq:weighted average ppe}
		&\mathcal{F}_{ph,ph}(N) = \nonumber\\
		&= \sum_{k=1}^{N} 
		\ev{\,\text{MS}(\varepsilon_k(\rho_{1,0})\otimes \rho_{2,0})_{real}\,}{\Phi}\,w_k \nonumber\\
		&= \frac{p(1-c)}{\left(1-(1-p)^{N}\right)}\ \times\ \frac{1-(1-c)^N(1-p)^N}{p+c\left(1 - p\right)}\ \times\nonumber\\
		&\ \ \ \times \frac{1}{60} \mathcal{F}_{2,0} \left(4\mathcal{F}_{1,0}-1\right) \left(16\mathcal{F}_{MS}-1\right) \nonumber\\
		&\ \ \  +\frac{1}{15} \left(1-\mathcal{F}_{MS}-\frac{1}{4}\mathcal{F}_{2,0}+4\mathcal{F}_{MS}\mathcal{F}_{2,0}\right)\ . 
	\end{align}
	This function is then used to fit the data of \autoref{sec: ExperimentPhPhResults} with $\mathcal{F}_{MS}$ and $c$ as fit parameters, the initial fidelities of \autoref{sec: ExperimentATPhResults}, and the generation and detection probability $p$ of an independent measurement.

	\section{Signal to background ratio}\label{appendix: SBR} The correlation of the 393\,nm laser pulse and the detected photons are evaluated for each atom and detector to infer the signal to background ratio (SBR) of the measurement. The wavepacket has a steep onset followed by an exponential decay which allows to use the time before the onset, i.e. before the 393\,nm excitation, to infer the background, and the sum over all events is used as signal. The result is listed in \autoref{tabSBR}.
	\begin{table}[h]
		\begin{minipage}{174pt}
			\caption{SBR for each detector.}\label{tabSBR}
			\centering 
			\begin{tabular}{l|r} 
				\toprule
				Detector & SBR \\ 
				\toprule
				atom\,1 detector\,1 & 810\\ 
				atom\,1 detector\,2 & 667\\
				atom\,2 detector\,1  & 322\\ 
				atom\,2 detector\,2  & 59\\ 
				\botrule
			\end{tabular}
		\end{minipage}
	\end{table}
	The SBR for atom\,1 is significantly higher than for atom\,2 which is still sufficiently large. This is mainly caused by the collection of background light on the path to the detectors.

	\section{M\o{}lmer-S\o{}rensen gate fidelity}\label{appendix: MSgate}
	The calibration of the M\o{}lmer-S\o{}rensen gate fidelity is accomplished by a parity measurement. Therefore the two atoms are prepared in $\ket{\text{S}_{1/2},m = +1/2}$ after Doppler cooling, then the M\o{}lmer-S\o{}rensen gate is applied on the axial sidebands of the $\ket{\text{S}_{1/2},m = +1/2}$-$\ket{+}$ transition. By scanning the phase of a global $\pi$/2-pulse and subsequent fluorescence detection, the amplitude $A$ and the projection result of $P_\text{DD+SS}$ (probability of both atoms in D$_{5/2}$ or S$_{1/2}$) is evaluated to reveal the fidelity
	\begin{align*}
		\mathcal{F}_{MS}=\frac{P_\text{DD+SS}}{2}+\frac{A}{2}\ .
	\end{align*}
	Evaluating all parity measurements that were done during the measurement of the photon-photon entanglement, a mean amplitude of $\bar{A}=0.89(2)$ and mean population of $\bar{P}_\text{DD+SS}=0.962(15)$ is calculated. This leads to a mean fidelity of $\bar{\mathcal{F}}_{MS}=0.926(17)$. 
	
	\section{Rate model to estimate superiority threshold}\label{appendix:rate_model}
	Different protocols are compared with regard to their realizable rates.
	The cases investigated are direct communication, semi-asynchronous (which is used in this paper) and fully-asynchronous protocol with the corresponding rates $r_{direct}, r_{s.asyn}$ and $r_{f.asyn}$.
	
	The rates are calculated by the probability for detection of photon pairs during the time $\tau$ which is the sum of the clock period $\tau_0$ and the communication time $\tau_C$. The communication time $\tau_C$ is the transmission time over the fiber from the central point to an end node (A or B according to \autoref{fig:seq_protocol}) and the time which is needed to send the detection information back.
	In the case of direct communication the source is located at A, so the photons must travel twice the distance as the photons of the cell.
	
	The other cases correspond to that of the discussed QR cell, where in the semi-asynchronous case, a photon is generated first on one side and then on the other.
	Finally in the fully-asynchronous case, photons are generated simultaneously on both sides until one is detected on each side.
	The formulas for the resulting rates in the different cases are
	\begin{align}
		r_{direct} &= \frac{p p_t^2}{\tau'}\ ,\\
		r_{s.asyn} &= \frac{0.5(1-(1-p p_t)^{n_\text{max}})^2}{n_\text{max}\tau}\ ,\\
		r_{s.asyn} &= \frac{0.5(1-(1-p p_t)^{n_\text{max}})^2}{n_\text{max}\tau}\ ,\\
		r_{f.asyn} &= \frac{(1-(1-p p_t)^{n_\text{max}})^2}{n_\text{max}\tau}
	\end{align}
	with the single-shot detection probability $p$ without additional transmission losses (same definition as in appendix \ref{appendix: ape_ppe_theo}), the transmission probability $p_t$ (e.g. through another fiber) and the maximum number of trials $n_\text{max}$. $\tau'$ indicates that in the synchronous case twice the fiber length as in the asynchronous cases must be considered.
	
	Exemplarily a case is selected, which was investigated in the experiment, viz. the one with $p=0.1\%$ and $p_t=24\%$, which corresponds to a fiber length from central station to one end node of 31.4\,km. As clock time the repetition time of step (4) in appendix \ref{appendix: protocol} of 4.5\,$\mu$s is used, which is clearly above the physical limit given by the linewidth of the transition which would result in a wavepacket of 741\,ns.
	
	For the selected case, there is a superiority of the fully-asynchronous protocol for $n_\text{max}\geq 573$ and for the semi-asynchronous one for $n_\text{max}\geq 1380$ with respect to direct communication.
	
	To avoid the need for this high number of trials, a higher detection probability $p$ must be achieved, this can be reached by the use of an objective with higher NA, instead of 0.4 one with 0.7. This would result in a 3.4 times higher detection probability $p=0.34\,\%$. This means that the limits are now lower, resulting in superiority when $n_\text{max}\geq 169$ and $n_\text{max}\geq 406$ for the fully-asynchronous and semi-asynchronous case, respectively.

\bibliography{main_arxiv}

\end{document}